\documentclass[aps,llncs, preprint]{revtex4-1} 
\usepackage{graphicx}
\usepackage{amsmath}
\usepackage{float}
\begin{document}
\title{The pressure-enhanced superconducting phase of Sr$_x$-Bi$_2$Se$_3$ probed by hard point contact spectroscopy} 

\author{Ritesh Kumar$^1$, Shekhar Das$^1$, Aastha Vasdev$^1$, Sandeep Howlader$^1$, Karn S. Jat$^2$, Prakriti Neha$^2$}

\author{Satyabrata Patnaik$^2$}
\author{Goutam Sheet$^1$}
\email{goutam@iisermohali.ac.in}
\affiliation{$^1$Department of Physical Sciences, 
Indian Institute of Science Education and Research Mohali, 
Mohali, Punjab, India}

\affiliation{$^2$School of Physical Sciences, Jawaharlal Nehru University, New Delhi, India}
 
\begin{abstract}

The superconducting systems emerging from topological insulators upon metal ion intercalation or application of high pressure are ideal for investigation of possible topological superconductivity. In this context, Sr-intercalated Bi$_2$Se$_3$ is specially interesting because it displays pressure induced re-entrant superconductivity where the high pressure phase shows almost two times higher $T_c$ than the ambient superconducting phase ( $T_C\sim$ 2.9 K). Interestingly, unlike the ambient phase, the pressure-induced superconducting phase shows strong indication of unconventional superconductivity. However, since the pressure-induced phase remains inaccessible to spectroscopic techniques, the detailed study of the phase remained an unattained goal. Here we show that the high-pressure phase can be realized under a mesoscopic  point contact, where transport spectroscopy can be used to probe the spectroscopic properties of the pressure-induced phase. We find that the point contact junctions on the high-pressure phase show unusual response to magnetic field supporting the possibility of unconventional superconductivity. 

\end{abstract}
\maketitle


  In superconductors, due to particle-hole symmetry, the positive and negative energy eigenstates of the Bogoliubov-DeGennes Hamiltonian  come in pairs\cite{Gennes,Alicea}. In the superconducting ground state, the negative-energy eigenstates are fully occupied. Therefore, as in case of insulators, depending on the dimension and the symmetries of the system, various topological numbers (e.g., the Chern number) for the occupied states can be defined\cite{Hasan,Kane,Fu,Qi}. If non-zero topological numbers exist for a superconductor, that can be classified as a "topological" superconductor \cite{Qi1,Schnyder,Qi2,Fu2}. By this definition, when certain unconventional superconductors display nodes in the order parameter symmetry, the node themselves might have non-zero topological numbers thereby making the superconductors "weakly" topological. On the other hand, in strong topological superconductors, the non-zero topological numbers can exist along with a fully gapped bulk superconducting gap. Hence, characterizing the topological nature of strong topological superconductors is a challenging task. However, due to topological restrictions, the surface of such superconductors host gap-less modes which can be detected by surface sensitive spectroscopic techniques\cite{Trang,Zhao,Sasaki,Xu,Wang}. Potentially, point-contact Andreev reflection can be a powerful technique to probe transport through such topological surface states in a topological superconductor\cite{Sasaki,Chen,Peng,Dai}. One popular route to possibly achieving topological superconductivity is doping charge carriers through metal intercalation in topological insulators like Bi$_2$Se$_3$\cite{Shruti,Hor,Liu,Han}. ARPES experiments have confirmed that at the required doping level for superconductivity ($\sim$ 2 $\times$ 10$^{20}$ cm$^{-}$$^{3}$) in charge doped Bi$_2$Se$_3$ systems, there is still significant separation in the momentum space between the topological surface states and the bulk states\cite{Han}. Hence, it is expected that when the bulk superconducting phase leads to proximity-induced superconductivity on the surface, due to the inherent topological nature of the surface states, the proximity induced phase should become a 2D topological superconductor\cite{Xu1,Wang,Fu2,Xu}. Another potentially interesting way of inducing superconductivity in a topological insulator is through applying pressure\cite{Zhang,Zhu,Zhou,Kirshenbaum}. A pressure-induced superconducting phase was indeed found in undoped Bi$_2$Se$_3$\cite{Kirshenbaum}. A more interesting pressure-induced superconducting phase was seen to appear in Sr-intercalated Bi$_2$Se$_3$ which shows ambient superconductivity below $T_c = $ 2.9 K\cite{Manikandan}. In this case, superconductivity was first seen to disappear with applying pressure and re-emerge at higher pressure\cite{Zhou}. The high-pressure re-entrant superconducting phase was found to be interesting owing to a significantly higher $T_c$ compared to the $T_c$ of the ambient superconducting phase of Sr-Bi$_2$Se$_3$. More importantly, the pressure-induced re-emerged phase showed strong signatures of unconventional superconductivity indicating a high possibility of the pressure-induced superconducting phase of Sr-Bi$_2$Se$_3$ being topological in nature. However, because technologically it is extremely challenging to perform spectroscopic investigation of the re-entrant phase, the exact nature of superconductivity in this phase remained poorly understood. In this paper, we discuss a unique way of realizing such a superconducting phase by applying uniaxial pressure under a point contact, where the superconducting phase can be investigated through mesoscopic transport spectroscopy.

\begin{figure}[h!]
		\includegraphics[width=.7\textwidth]{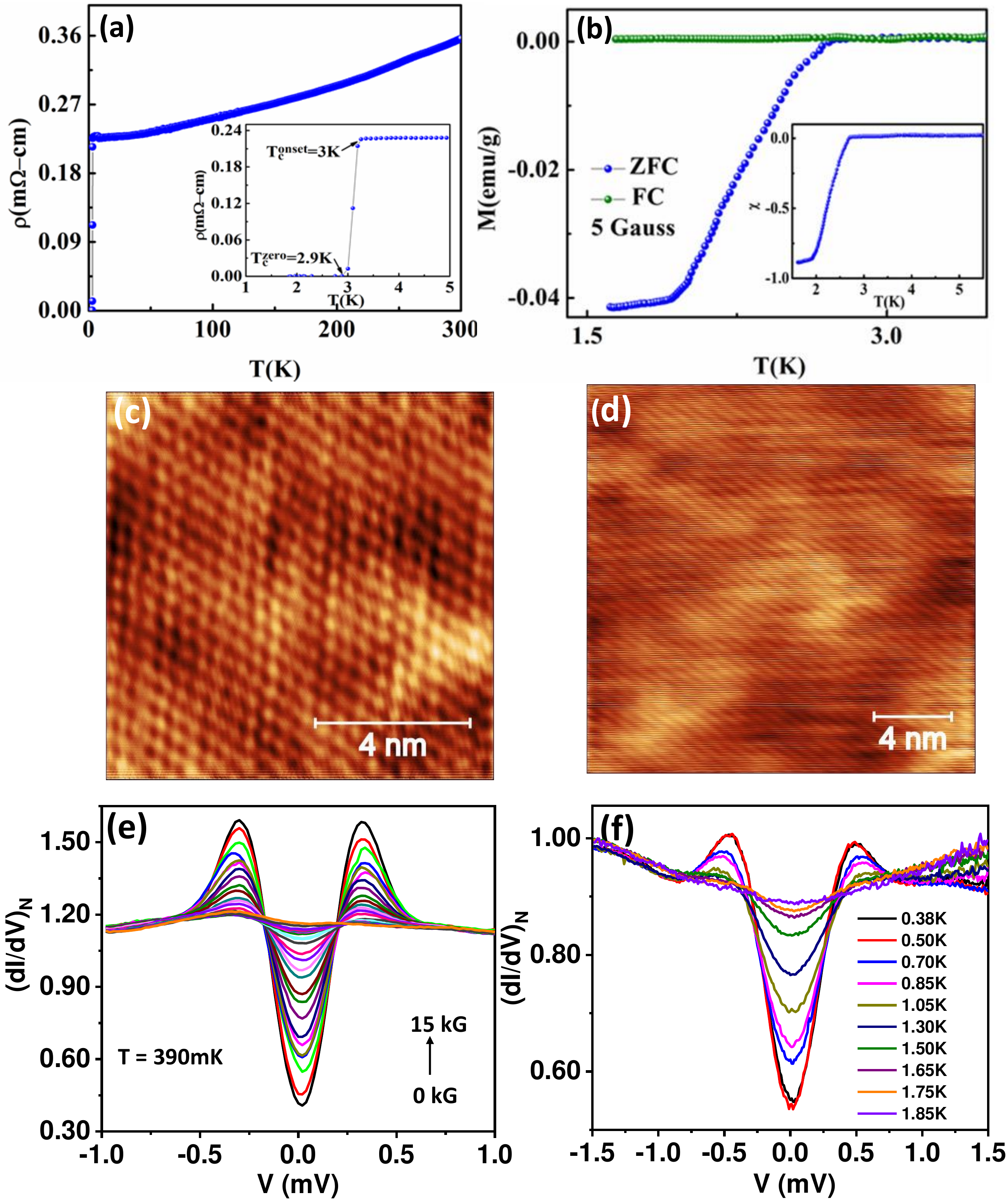}
		\caption{(a) Resistivity ($\rho$) as a function of temperature ($T$) of Sr$_{0.1}$Bi$_2$Se$_3$ (b) Magnetic susceptibility vs temperature (zero field cooled (ZFC) and Field Cooled (FC)). (c) Atomic resolution image (10 nm X 10 nm) of the Sr$_{0.1}$Bi$_2$Se$_3$ surface. (d) STM topography of the Sr$_{0.1}$Bi$_2$Se$_3$ cleaved surface (20 nm X 20 nm). (e) Normalized STS spectra with varying magnetic fields upto 15 kG. (f) Temperature dependence of the STS spectra upto 1.85 K.}	

\end{figure}

We have performed experiments on high quality single crystals of Sr$_{0.1}$Bi$_2$Se$_3$. The bulk magnetization (Figure 1(a)) and transport measurements (Figure 1(b)) revealed a critical temperature $T_c\sim$ 2.9 K below which the system superconducts. The high quality of the crystals were further confirmed by atomic resolution scanning tunnelling microscopy and spectroscopy. As shown in Figure 1(c), the atomic lattice is seen with very low defect density. Tunnelling spectroscopy revealed a fully formed superconducting gap that evolves systematically with increasing temperature and near 2 K the spectra become too broad for the gap to be clearly seen. The gap also evolves systematically with magnetic field before being almost completely suppressed at 15 kG (Figure 1(e)).   
\begin{figure}[h!]
		\includegraphics[width=1\textwidth]{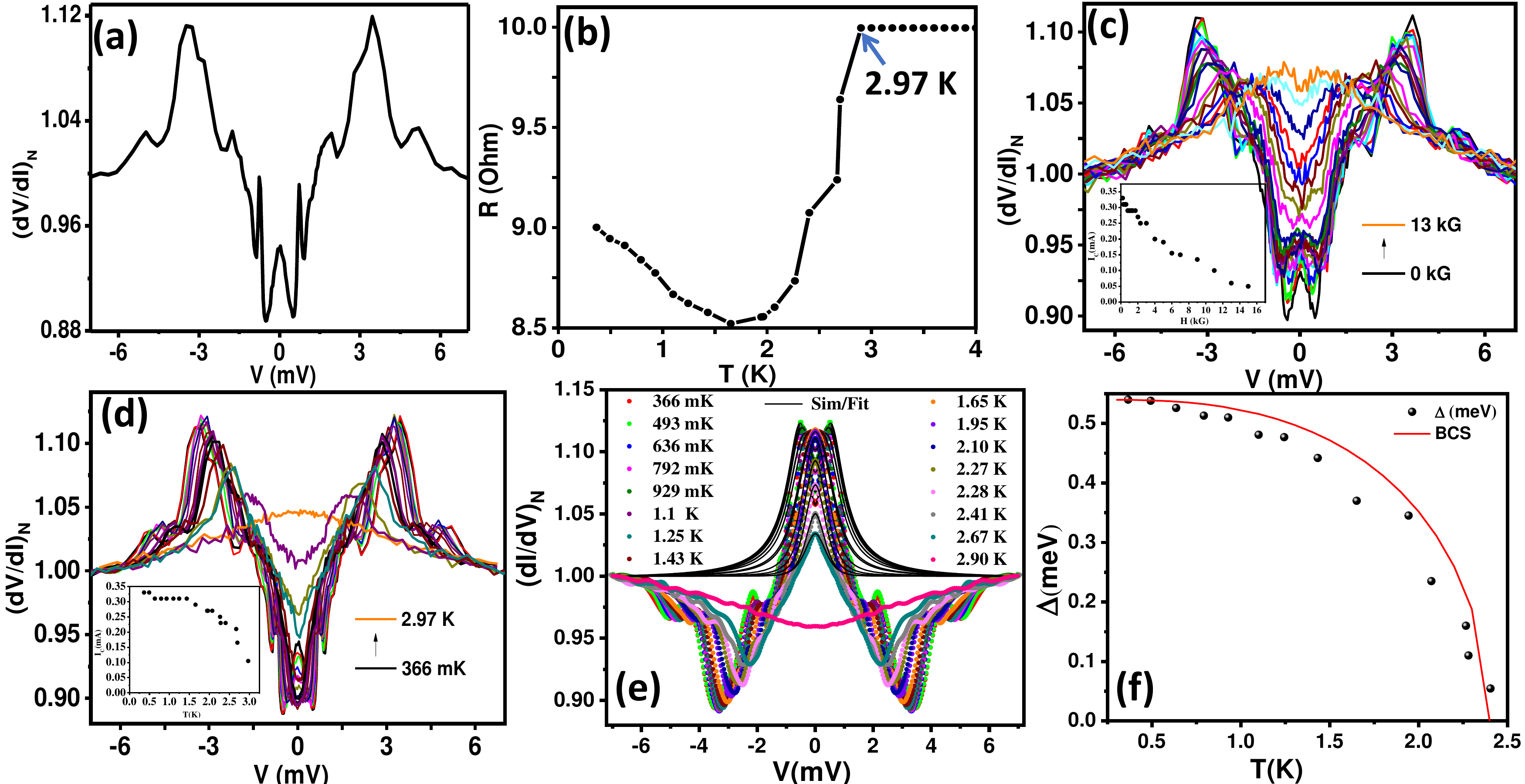}
        \caption{(a) Normalized ($(dV/dI)_N$) spectrum obtained in the intermediate regime of transport showing the characteristic signatures of critical current peaks and Andreev refection (AR) dips. (b) Zero bias resistance ($R$) vs. $T$ of the point contact in intermediate regime. (c,d) Magnetic field ($H$) and temperature dependence of the spectrum shown in (a). The $inset$ shows magnetic field and temperature dependence of the current corresponding to the peaks ($I_c$) extracted from the data in Figure 2(c) and 2(d) respectively. (e) Temperature dependence of the Intermediate limit spectra (coloured dots) along with the best theoretical BTK fits (solid black lines). (f) $\Delta$ vs. $T$ plot extracted from (e). The Red line shows the expected BCS line.}	

\end{figure}

These observations for the ambient superconducting phase of Sr-Bi$_2$Se$_3$ are consistent with the previous experiments\cite{Manikandan}. Here our aim was to probe the pressure-induced re-entrant, enhanced superconducting phase of Sr-Bi$_2$Se$_3$. For that we used a home-built point contact spectroscopy set up working down to 400 mK\cite{Das}. First a silver (Ag) tip was engaged gently on the surface of the crystal thus forming mesoscopic point contact junctions between Ag and Sr-Bi$_2$Se$_3$. The smallest area point contacts revealed dips in $dV/dI$ vs. $V$ spectra (Figure 2(a)) showing signature of Andreev reflection. Such spectra also contained multiple peaks in $dV/dI$ indicating that the point-contacts being away from the ballistic regime of transport where the peaks appear due to critical current dominated non-linearities in $I-V$. As shown in Figure 2(b), temperature dependent measurement of the zero-bias resistance of the point contact revealed a superconducting transition near 2.9 K confirming that the point contact was probing the ambient superconducting phase and the pressure under the point contact was still very small.

\begin{figure}[h!]
		\includegraphics[width=1\textwidth]{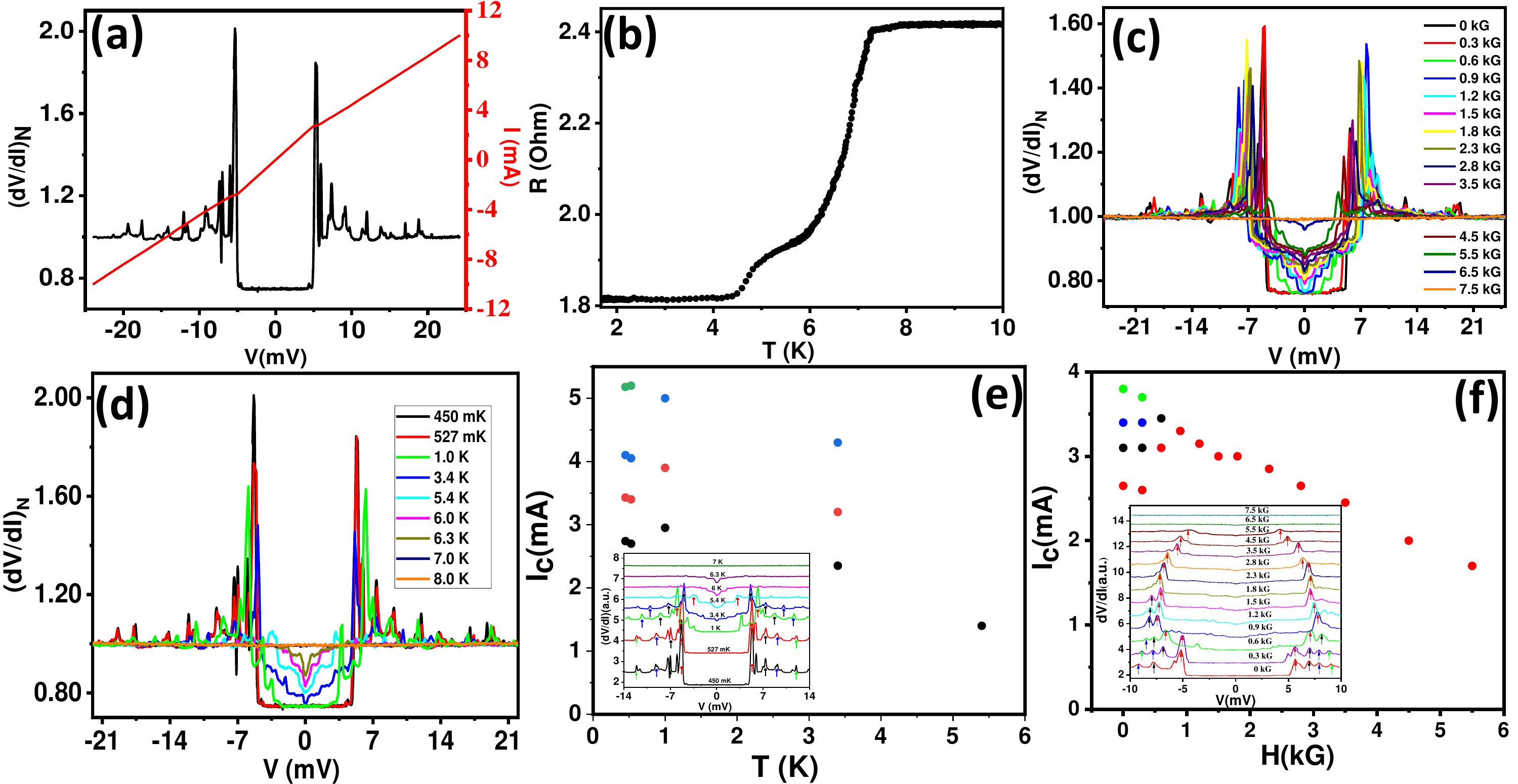}
		\caption{(a) Normalized differential resistance ($(dV/dI)_N$) spectrum in the thermal regime showing the multiple critical current peaks. The red line shows the $I–V$ plot corresponding to the total resistance of the point contact. (b) Resistance ($R$) of the point contact vs. $T$ in thermal regime. (c,d) Magnetic field and temperature dependence of the spectrum shown in (a). (e) Temperature dependence the critical current corresponding to the peaks structure ($I_c$) with $T$ extracted from the data in (d). Inset shows the critical current peak position marking with coloured arrows. (f) Magnetic field dependence of the critical current corresponding to the peaks structure ($I_c$) with $H$ extracted from the data in (c). The $inset$ shows the multiple critical current peaks (marked with coloured arrows).}	

\end{figure}

After that, we continued pressing the tip harder onto the crystal surface. During the process, we found signature of superconductivity at a temperature higher than the $T_c$ of pristine Sr-Bi$_2$Se$_3$. As it is seen in Figure 3(a), upon applying large pressure under the point contact, superconductivity at a higher temperature is achieved, but now due to large force applied, the contact diameter also changed thereby forming a point contact in the thermal regime of transport. In this extreme thermal regime, features associated with Andreev reflection completely disappeared and multiple sharp peaks in $dV/dI$ appeared indicating formation of multiple thermal point contacts under the contact region. A temperature dependent measurement of the point contact resistance revealed a critical temperature $T_c\sim$ 8 K which corresponds to the known pressure-induced re-entrant superconducting phase of Sr-Bi$_2$Se$_3$\cite{Zhou}. As it is seen in the point contact $R-T$ data, the transition is broad and at a relatively lower temperature (around 6 K), another transition like feature is seen. These could be attributed to multiple electrical contacts with different contact geometries formed. Each contact may experience different pressure due to the difference in their effective contact area. Comparing the measured $T_c$ with the published literature\cite{Zhou}, we estimate the approximate pressure experienced by the superconducting region under the point contact to be 9 GPa.

In order to gain further understanding on the pressure-enhanced superconducting phase, we carried out detailed temperature and magnetic field dependent experiments. As seen in Figure 3(d), the point contact spectrum evolves monotonically with temperature. At the lowest temperature (450 mK), the critical-current dominated features (peaks in $dV/dI$) are extremely sharp. With increasing temperature, all the peaks shift and they come closer indicating temperature dependent suppression of critical current for each micro-constriction formed under the point contact. Finally, at 8 K, all the features associated with superconductivity disappear. The critical current driven peaks also shift inward with increasing magnetic field (Figure 3(f)). For all the micro-constrictions, the critical current is seen to decrease at a slow rate. For the constriction with highest critical current (red dots in Figure 3(f)), the critical current shows slight increase at lower fields and then starts decreasing slowly. At a field of 6 kG, the critical current has become only half of the zero field value. The over-all superconductivity-related spectral features completely disappear at 7.5 kG. 

In order to find out whether the unusual magnetic field dependence is also seen in the transport experiments, we have analyzed the $R$ vs. $T$ data of the thermal limit point contact obtained at different magnetic fields. The field-dependent $R-T$ curves are shown in Figure 4(a). We have tracked the shift in transition at higher temperature with magnetic field to construct the $H-T$ phase diagram. As shown in  Figure 4(b), the experimentally obtained $H-T$ data shows dramatic deviation from the conventional $H-T$ curve that is usually seen in conventional superconductors. These observations indicate the possibility of an unconventional component in the superconducting order parameter of the high-pressure phase\cite{Zhou}. 

It should be noted that despite multiple attempts, a ballistic point contact could not be realized in this phase as during our efforts to reduce the contact diameter through controlled withdrawal of the tip, the effective pressure also decreased thereby causing a sudden disappearance of the pressure-induced phase.

\begin{figure}[h!]
		\includegraphics[width=1\textwidth]{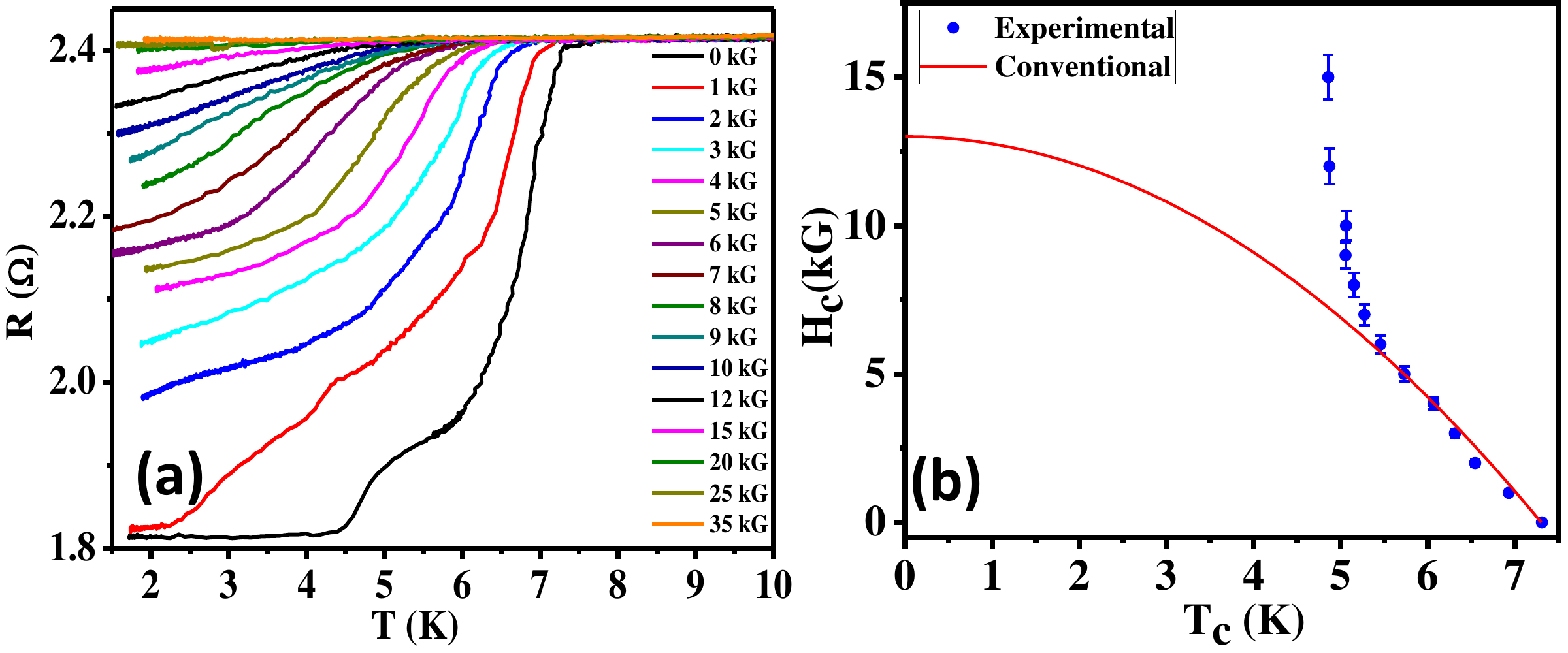}
		\caption{(a) Magnetic field dependence of the $R–T$ curves. (b) $H–T$ phase diagram. The red line shows the empirical plot for a conventional superconductor. The blue dots are the data points extracted from the graphs in (a).}	

\end{figure}

Therefore, we have realized the pressure-enhanced superconducting phase of Sr-intercalated Bi$_2$Se$_3$ under hard point contacts and investigated both the low-pressure and the high-pressure phases spectroscopically. We found that while the low-pressure superconducting phase behaves like a conventional superconductor, the high-pressure phase has unusual magnetic properties. The critical current of the thermal point contacts formed with the high-pressure superconducting phase shows unusual rigidity with increasing magnetic field. Furthermore, The $H-T$ phase diagram of the high-pressure phase shows dramatic deviation from a conventional convex shape. Such observations indicate the possibility of unconventional superconductivity (and, topological superconductivity) in the high-pressure superconducting phase of Sr-intercalated Bi$_2$Se$_3$. This work also demonstrates an unique way of spectroscopically probing pressure-induced or pressure-enhanced superconducting phases in new generation quantum materials.   

GS acknowledges financial support from Swarnajayanti fellowship awarded by the Department of Science and Technology (DST), Govt. of India (grant number DST/SJF/PSA-01/2015-16).

\end{document}